\documentclass[sigconf]{acmart}

\usepackage{booktabs} 
\usepackage{subcaption}
\usepackage{balance}
\setcopyright{acmcopyright}
  \usepackage{etoolbox}
 \usepackage{mathrsfs}
 \usepackage{mathtools}

\def\BibTeX{{\rm B\kern-.05em{\sc i\kern-.025em b}\kern-.08emT\kern-.1667em\lower.7ex\hbox{E}\kern-.125emX}}

\copyrightyear{2019} 
\acmYear{2019} 
\acmConference[CIKM '19]{The 28th ACM International Conference on Information and Knowledge Management}{November 3--7, 2019}{Beijing, China}
\acmBooktitle{The 28th ACM International Conference on Information and Knowledge Management (CIKM '19), November 3--7, 2019, Beijing, China}
\acmPrice{15.00}
\acmDOI{10.1145/3357384.3358095}
\acmISBN{978-1-4503-6976-3/19/11}

\begin{document}
\title{A Study of Context Dependencies in Multi-page Product Search}

\author{Keping Bi$^1$, Choon Hui Teo$^2$,  Yesh Dattatreya$^2$, Vijai Mohan$^2$, W. Bruce Croft$^1$}
\affiliation{%
	\institution{$^1$College of Information and Computer Sciences, University of Massachusetts Amherst, Amherst, MA, USA}
}
\email{{kbi, croft}@cs.umass.edu}
\affiliation{%
	\institution{$^2$Search Labs, Amazon, Palo Alto, CA, USA}
}
\email{{choonhui, ydatta, vijaim}@amazon.com}

\begin{abstract}
In product search, users tend to browse results on multiple search result pages (SERPs) (e.g., for queries on clothing and shoes) before deciding which item to purchase. Users' clicks can be considered as implicit feedback which indicates their preferences and used to re-rank subsequent SERPs. Relevance feedback (RF) techniques are usually involved to deal with such scenarios. However, these methods are designed for document retrieval, where relevance is the most important criterion. In contrast, product search engines need to retrieve items that are not only relevant but also satisfactory in terms of customers' preferences. Personalization based on users' purchase history has been shown to be effective in product search \cite{ai2017learning}. However, this method captures users' long-term interest, which do not always align with their short-term interest, and does not benefit customers with little or no purchase history. 
In this paper, we study RF techniques based on both long-term and short-term context dependencies in multi-page product search. We also propose an end-to-end context-aware embedding model which can capture both types of context. 
Our experimental results show that short-term context leads to much better performance compared with long-term and no context. Moreover, our proposed model is more effective than state-of-art word-based RF models.

\end{abstract}

%
%

\keywords{Implicit Feedback; Multi-page Product Search; Context-aware Search}

\fancyhead{}
\maketitle

\section{Introduction}
\label{sec:introduction}

Users tend to browse multiple SERPs to view more products and make comparisons before they make final purchase decisions in product search. From the search log of a retailer's online store, we observe that in about 5\% to 15\% of search traffic, users browse and click results in the previous pages and purchase items in the later result pages. Users' clicks can be considered as implicit feedback that indicates their preference in the current query session. Relevance feedback (RF) approaches can be used to extract the relevance topic and re-rank the subsequent SERPs. There has been some research on multi-page search \cite{jin2013interactive,zeng2018multi}.
However, these methods are designed for document retrieval, which has different characteristics from product search. 
Documents consist of text while products are essentially entities that have many aspects such as brand, color, size and so on. In addition, 
in contrast to document retrieval, where relevance is a universal evaluation criterion, a product search system is evaluated based on user purchases that depend on both product relevance and customer preferences. In this paper, we study the problem of multi-page product search, where little research has been conducted. 

Most previous studies on product search focus on product relevance \cite{duan2013probabilistic, van2016learning, wu2018turning, karmaker2017application}. 
Attempts were also made to improve customer satisfaction by diversifying search results \cite{yu2014latent}. 
Recently, \citet{ai2017learning} introduced a personalized ranking model which takes the users' preferences learned from their historical reviews together with the queries as the basis for ranking. However, the personalized model cannot cope with the situations such as users that have not logged in during searching and thus can not be identified, users that logged in but do not have enough purchase history, and a single account being shared by several family members. In these cases, user purchase records are either not available or containing substantial noise. 
Moreover, users' long-term behaviors may not be as informative to indicate the user's preferences as short-term behaviors such as interactions with the shown items in a query session. These limitations of existing work on product search motivate us to study short-term feedback for modeling user preferences in a query session, which do not require additional customers' information or their purchase history, and compare long-term and short-term context in multi-page product search. 

Traditional relevance feedback (RF) methods, which extract expansion terms from feedback documents, have potential word mismatch problems \cite{Zamani:2016:EQL:2970398.2970405}. To tackle this problem, we propose an end-to-end context-aware embedding model that can incorporate both long-term and short-term context to predict purchased items. In this way, semantic match and the co-occurence relationship between clicked and purchased items are both captured in the embeddings. 
We show the effectiveness of incorporating short-term context against baselines using both long-term context and no context. Also, our model performs better than state-of-art word-based RF models by a large margin.  

Our contributions can be summarized as follows: (1) we reformulate conventional one-shot ranking to dynamic ranking (i.e., multi-page search) based on user clicks in product search; (2) we introduce different context dependency assumptions and propose a simple yet effective end-to-end embedding model to capture different types of dependency; (3) we investigate different aspects in multi-page product search on real search log data and show the effectiveness of incorporating short-term context and neural embeddings. Our study on multi-page product search indicates that this is a promising direction and worth more attention. 

\section{Related Work}
\label{sec:related_work}
\textbf{Product Search.}
\label{subsec:product_search}
Most previous work treats product search as a one-shot ranking problem, where given a query, static results are shown to users regardless of their interaction with the result lists. Facets of products have been used for product search ~\cite{lim2010multi, vandic2013facet}. Language model based approaches have been studied to support keyword search ~\cite{duan2013probabilistic}. Later, to further solve vocabulary mismatch, models that measure semantic match between queries and products based on reviews have been proposed \cite{van2016learning, ai2017learning}. Other aspects of product search such as popularity, visual preference and diversity have also been studied \cite{long2012enhancing, guo2018multi, yu2014latent}. In terms of labels for training, there are studies on using clicks, purchases, click-rate, add-to-cart ratios and order rates as labels \cite{wu2018turning, karmaker2017application}. 
In a different approach, \citet{hu2018reinforcement} use online reinforcement learning mechanism to rank products dynamically when users request next SERPs. However, they update a global ranker given the signal of purchases. In contrast, our model updates SERPs for each individual query based on the clicks collected under the query. 

\textbf{Multi-page Search and Relevance Feedback (RF). }
Some research has been conducted on multi-page search \cite{jin2013interactive,zeng2018multi}. They are word-based or learning-to-rank based methods and focus on document retrieval where relevance plays a different role than in product search. 
There has been considerable research on RF. However, most of them are unsupervised methods and based on bag-of-word representations, such as Rocchio \cite{rocchio1971relevance} and the Relevance Model (RM3) \cite{lavrenko2017relevance}. Embedding-based RF methods have also been proposed to leverage semantic match \cite{Zamani:2016:EQL:2970398.2970405, DBLP:conf/ecir/BiAC19}. 
Although these RF methods can also be applied in our task, we propose an end-to-end neural model for RF in the context of product search.

\section{Context-aware Product Search}
\label{sec:context_ps}
We first formulate the task of multi-page product search. Then we discuss different assumptions of context dependency models and propose a context-aware embedding model for the task.


\subsection{Problem Formulation}
\label{subsec:prob_form}
A query session\footnote{We refer to the series of user behaviors associated with a query as  a query session, i.e, a user issues a query, clicks results, paginates, purchases items and finally ends searching with the query. } is initiated when a user $u$ issues a query $q$ to the search engine. 
Let $R_t$ be the set of items on the $t$-th search result page ranked by an initial ranker and denote by $R_{1:t}$ the union of $R_1, \cdots, R_t$. For practical purposes, we let the re-ranking candidate set $D_{t+1}$ for page $t+1$ be $R_{1:t+k}  \diagdown V_{1:t}$ where $k \geq 1$ and $V_{1:t}$ is the set of re-ranked items viewed by the user in the first $t$ pages. Given user $u$, query $q$, and the set of clicked items in the first $t$ pages $C_{1:t}$ as context, the objective is to rank all, if any, purchased items $B_{t+1}$ in $D_{t+1}$ at the top of the next result page. 


\subsection{Context Dependency Models}
\label{subsec:depend_context}
\begin{figure}
	\includegraphics[width=0.45\textwidth]{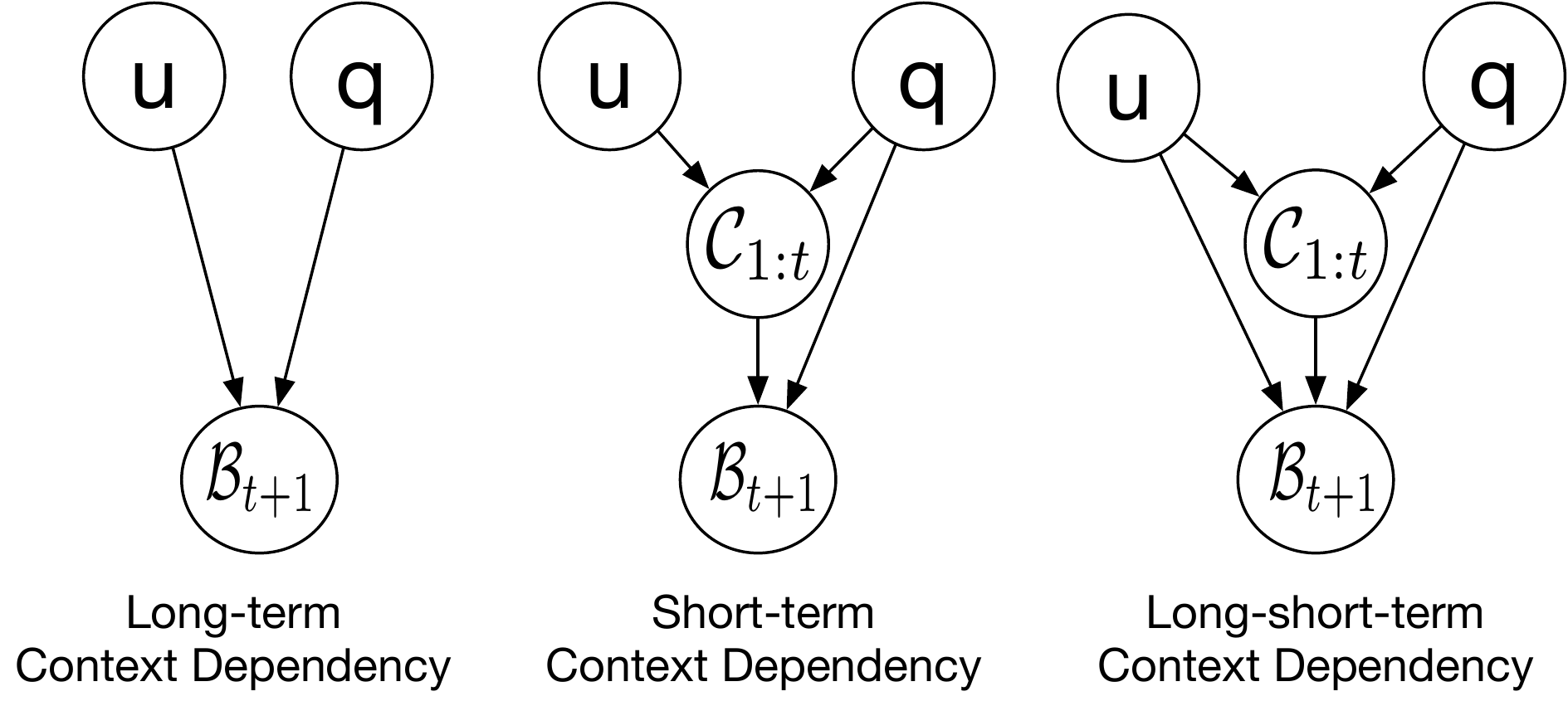} %
	\caption{Different assumptions to model different factors as context for purchase prediction.}
	\label{fig:depend}
\end{figure}

Figure \ref{fig:depend} shows the graphical models for three types of context dependencies, long-term context, short-term context, and long-short-term context. $u$ denotes the latent variable of a user's long-term interest independent of queries, and clicks in the first $t$ result pages, i.e., $C_{1:t}$, represents the user's short-term preference. Purchased items on and after page $t+1$, i.e., $\mathcal{B}_{t+1}$, depends on query $q$ and different types of context under different dependency assumptions.


\textbf{Long-term Context Dependency.} Only users' long-term preferences, usually represented by their historical queries and the corresponding purchased items (denoted as $u$ in Figure \ref{fig:depend}) are used to predict purchased items. Such models can provide personalized search results at the beginning of a query session \cite{ai2017learning}. However, this assumption needs user identity and purchase history, which are not always available. Moreover, it may not be informative to predict the final purchase since users' current search intent may be different from any of her previous searches and purchases. 

\textbf{Short-term Context Dependency. }
In this assumption, given the observed clicks in the first $t$ pages ($\mathcal{C}_{1:t}$) as short-term context, the items purchased in the subsequent result pages ($\mathcal{B}_{t+1}$), are conditionally independent of the user $u$, shown in Figure \ref{fig:depend}. 
Users with little or no purchase history and who have not logged in can benefit directly under such a ranking scheme.

\textbf{Long-short-term Context Dependency.}
In this model, an unseen item $i$ after page $t$ is scored according to $p(i \in \mathcal{B}_{t+1} | \mathcal{C}_{1:t}, q, u)$, which considers both long-term context ($u$) and short-term context ($\mathcal{C}_{1:t}$).
This setting considers more information but it also has the drawback of requiring users identity and purchase history. 

In this paper, we focus on non-personalized short-term context and include the other two types of context for comparison.

\subsection{Context-aware Embedding Model}
\label{subsec:embed_ca_model}
\begin{figure}
	\includegraphics[width= 0.45 \textwidth]{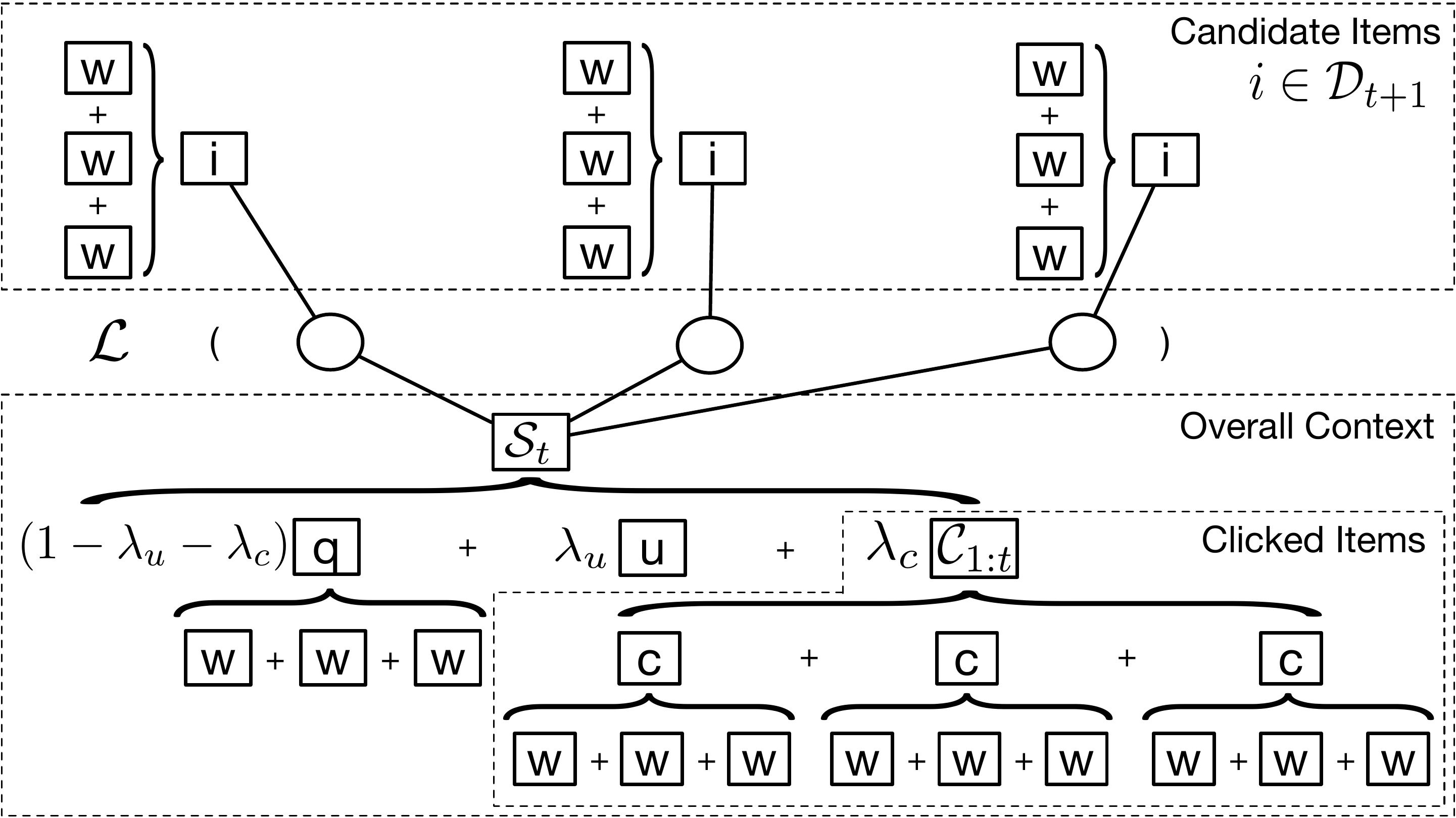} %
	\caption{Our context-aware embedding model (CEM).  }
	\label{fig:model}
\end{figure}

We designed a context-aware model (CEM) where different dependency assumptions can be captured by varying the corresponding coefficients, shown in Figure \ref{fig:model}. 

\textbf{Item Embeddings.}
We use product titles to represent products since merchants tend to put the most informative, representative text such as the brand, size, color, material and even target customers in product titles. We use the average of title word embeddings of a product as its own embedding ($\mathcal{E}(i)$). 
\footnote{Other encoding methods such as non-linear projection of average word embeddings and recurrent neural network have not performed better than this simple method.} 
In this way, word representations can be generalized to new items, and we do not need to cope with the cold-start problem.

\textbf{User Embeddings.}
Each user has a unique representation $\mathcal{E}(u)$ from a lookup table, which is shared across search sessions and updated by the gradient learned from previous user transactions. In this way, the long-term interest of the user is captured and we use the user embeddings as long-term context in our models.

\textbf{Query Embeddings.}
Similar to item embeddings, we use the average embedding of query words as the representation ($\mathcal{E}(q)$).

\textbf{Short-term Context Embeddings. }
We use the set of clicked items to represent user preference behind the query. 
We assume the sequence of clicked item does not matter when modeling short-term user preference under the same query. 
One reason is that user's purchase needs are often fixed given her query. Another reason is that the order of user clicks is usually based on the rank of retrieved products as users examine each result from top to bottom. 
Similar to \cite{rocchio1971relevance, DBLP:conf/ecir/BiAC19}, we represent the relevant set with the centroid of each item in the set and average item embeddings are used to represent the centroid, denoted as $\mathcal{E}(\mathcal{C}_{1:t})$. 
\footnote{We also tried an attention mechanism to weight each clicked item according to the query and represent the user preference with a weighted combination of clicked items. However, this method is not better than combining clicks with equal weights. }


\textbf{Overall Context Embeddings.}
We use a convex combination of user, query, and click embeddings as the representation of overall context $\mathcal{E}(\mathcal{S}_t)$. i.e.
\begin{equation}
\label{eq:context}
\begin{aligned}
\mathcal{E}(\mathcal{S}_t) &= (1-\lambda_u-\lambda_c) \mathcal{E}(q) + \lambda_u \mathcal{E}(u) + \lambda_c \mathcal{E}(\mathcal{C}_{1:t}) \\
& 0 \leq \lambda_u \leq 1, 0 \leq \lambda_c \leq 1, \lambda_u + \lambda_c \leq 1 \\
\end{aligned}
\end{equation}
This overall context is then treated as the basis for predicting purchased items in $\mathcal{B}_{t+1}$. 
When $\lambda_c$ or $\lambda_u$ is set to 0, the corresponding short-term or long-term context does not take effect. In other cases, both types of context are considered. By varying the values of $\lambda_u$ and $\lambda_c$, we can use Equation \ref{eq:context} to model different types of context dependency and do comparisons. 

\textbf{Attention Allocation Model for Items. } 
With the overall context collected from the first $t$ pages, we further construct an attentive model to re-rank the products in the candidate set $\mathcal{D}_{t+1}$. This re-ranking process can be considered as an attention allocation problem. Given the context that indicates the user's preference and a set of candidate items that have not been shown to the users yet, the item which attracts more user attention will have higher probability to be purchased. The attention weights then act as the basis for re-ranking. They can be computed as:

\begin{equation}
\label{eq:attn_score}
\begin{aligned}
score(i | q,u,\mathcal{C}_{1:t}) = \frac{\exp(\mathcal{E}(\mathcal{S}_t) \cdot \mathcal{E}(i))}
{\sum_{i' \in \mathcal{D}_{t+1}}\exp( \mathcal{E}(\mathcal{S}_t) \cdot \mathcal{E}(i'))} \\
\end{aligned}
\end{equation}
where $\mathcal{E}(\mathcal{S}_t)$ is computed according to Equation \ref{eq:context}. This function can also be interpreted as the generative probability of an item in the candidate set $\mathcal{D}_{t+1}$ given the context $\mathcal{S}_{t}$. 
Then the model is trained by maximizing the likelihood of observing $\mathcal{B}_{t+1}$ conditioning on corresponding $\mathcal{C}_{1:t}, u, q$ in the training set. 

\section{Experiments}
\label{sec:exp_setup}

\begin{table*}
	\caption{Performance of our short-term context embedding model (SCEM) and baselines when re-ranking from the 2nd page. 
		The number is the relative improvement of each method compared with the production model (PROD)\protect\footnotemark. `$^-$' indicates significant worse than SCEM in paired student t-test with $p \leq 0.001$.
		Note that difference larger than 3\% is approximately significant.} 
	\label{tab:overallperf}
	\scalebox{0.88}{    
		\begin{tabular}{  p{1.5cm} || r | r | r || r | r | r || r | r | r   }
			\hline
			& \multicolumn{3}{c||}{Toys \& Games} & \multicolumn{3}{c||}{Garden \& Outdoor} & \multicolumn{3}{c}{Cell Phones \& Accessories} \\
			\hline
			Model & $MAP$ & $MRR$ & $NDCG@10$ & $MAP$ & $MRR$ & $NDCG@10$ & $MAP$ & $MRR$ & $NDCG@10$ \\
			\hline
			\hline
			PROD & 0.00\%$^{-}$ & 0.00\%$^{-}$ & 0.00\%$^{-}$ & 0.00\%$^{-}$ & 0.00\%$^{-}$ & 0.00\%$^{-}$ & 0.00\%$^{-}$ & 0.00\%$^{-}$ & 0.00\%$^{-}$ \\
			\hline
			RAND & -25.70\%$^{-}$ & -26.83\%$^{-}$ & -29.23\%$^{-}$ & -23.40\%$^{-}$ & -24.16\%$^{-}$ & -25.73\%$^{-}$ & -20.15\%$^{-}$ & -20.93\%$^{-}$ & -22.73\%$^{-}$ \\
			\hline
			POP & -15.82\%$^{-}$ & -15.90\%$^{-}$ & -17.87\%$^{-}$ & -9.38\%$^{-}$ & -9.51\%$^{-}$ & -9.55\%$^{-}$ & -8.54\%$^{-}$ & -8.25\%$^{-}$ & -11.12\%$^{-}$ \\
			\hline
			QL & -25.78\%$^{-}$ & -27.80\%$^{-}$ & -29.73\%$^{-}$ & -19.62\%$^{-}$ & -20.78\%$^{-}$ & -21.63\%$^{-}$ & -16.14\%$^{-}$ & -16.77\%$^{-}$ & -18.00\%$^{-}$ \\
			\hline
			QEM & -2.57\%$^{-}$ & -3.10\%$^{-}$ & -3.85\%$^{-}$ & +0.65\%$^{-}$ & -0.34\%$^{-}$ & +1.06\%$^{-}$ & +9.96\%$^{-}$ & +9.73\%$^{-}$ & +10.58\%$^{-}$ \\
			\hline
			\hline
			LCRM3 & -24.82\%$^{-}$ & -25.92\%$^{-}$ & -28.60\%$^{-}$ & -19.33\%$^{-}$ & -20.45\%$^{-}$ & -21.28\%$^{-}$ & -15.44\%$^{-}$ & -16.07\%$^{-}$ & -17.38\%$^{-}$ \\
			\hline
			LCEM & -2.57\%$^{-}$ & -3.10\%$^{-}$ & -3.85\%$^{-}$ & +0.65\%$^{-}$ & -0.34\%$^{-}$ & +1.06\%$^{-}$ & +9.96\%$^{-}$ & +9.73\%$^{-}$ & +10.58\%$^{-}$ \\
			\hline
			\hline
			SCRM3 & +12.93\%$^{-}$ & +9.63\%$^{-}$ & +9.53\%$^{-}$ & +25.15\%$^{-}$ & +23.01\%$^{-}$ & +23.15\%$^{-}$ & +18.65\%$^{-}$ & +16.77\%$^{-}$ & +17.11\%$^{-}$ \\
			\hline
			SCEM & \textbf{+26.59\%} & \textbf{+24.56\%} & \textbf{+26.20\%} & \textbf{+37.43\%} & \textbf{+35.16\%} & \textbf{+37.22\%} & \textbf{+48.99\%} & \textbf{+47.00\%} & \textbf{+50.18\%} \\
			\hline
			\hline
			LSCEM & \textbf{+26.59\%} & \textbf{+24.56\%} & \textbf{+26.20\%} & \textbf{+37.43\%} & \textbf{+35.16\%} & \textbf{+37.22\%} & \textbf{+48.99\%} & \textbf{+47.00\%} & \textbf{+50.18\%} \\
			\hline
			\hline
		\end{tabular}
	}
\end{table*}

\textbf{Datasets}. We randomly sampled three category-specific datasets, namely, ``Toys \& Games'', ``Garden \& Outdoor'', and  ``Cell Phones \& Accessories'', from the logs of a commercial product search engine spanning ten months between 2017 and 2018. 
We keep only the query sessions with at least one clicked item on any page before the pages with purchased items. 
Our datasets include up to a few million query sessions containing several hundred thousand unique queries.
The average lengths of product titles in these categories are from 13 to 22 and vocabulary sizes are from 0.2M to 1M.  

\textbf{Evaluation Methodology.} 
The sessions that occurred in the first 34 weeks are used for training, the following 2 weeks for validation and the last 4 weeks for testing. 
Given that the datasets are static, we can only evaluate the performance of one-shot re-ranking from page $t+1$ given the context collected from the first $t$ pages. We experimented on the cases when $t=1$ \footnote{We also experimented the setting of $t=2$, results show similar trends and the improvements are larger since there are more clicks available in the first two SERPs.}. 
As in \cite{rocchio1971relevance}, only rank lists of unseen items are evaluated. $MAP$ at cutoff 100, $MRR$ and $NDCG$ at 10 are used as the metrics. 

\textbf{Baselines.}
We compare our short-term context-aware embedding model (\textit{SCEM}) with four groups of baseline: retrieval models without using context, long-term, short-term and long-short-term context-aware models. 
Baselines without using context include the production model (\textit{PROD}) (a state-of-art learning-to-rank model) and models that re-rank the results in the candidate set retrieved by PROD by randomly shuffle (\textit{RAND}), popularity (\textit{POP}), the query likelihood model (\textit{QL}) \cite{ponte1998language}, or the query embedding based model (\textit{QEM}) (CEM with $\lambda_u = 0, \lambda_c = 0$). 
Long-term context-aware baselines are the relevance model (RM3) \cite{lavrenko2017relevance} applied on the titles of the user's historical purchased products, denoted as \textit{LCRM3}, and long-term context-aware embedding model (\textit{LCEM}), which sets $\lambda_c = 0, 0 \leq \lambda_u \leq 1$ in CEM. 
RM3 that considers the clicked items in the current query session as positive feedback serves as the short-term context-aware baseline, denoted as \textit{SCRM3}. 
\footnote{We also tested the embedding-based relevance model (ERM) \cite{Zamani:2016:EQL:2970398.2970405}
	as an embedding-based baseline. However, it does not perform better than RM3 across different settings, so it was not included.  }
When both long-term and short-term context are incorporated in CEM, i.e., $\lambda_u \geq 0, \lambda_c \geq 0$ in Equation \ref{eq:context}, the model is referred to as long-short-term context-aware embedding model (\textit{LSCEM}). 

\textbf{Training.}
Query sessions with multiple purchases on different pages were split into sub-sessions, one for each page with a purchase. 
We trained our models with Tensorflow for 20 epochs with 256 samples in each batch.
Based on our validation results, we set $\lambda_c$ to 1 for SCRM3, SCEM, and LSCEM; $\lambda_u$ was set to 0.8 for LCRM3 and 1 for LCEM. 

\textbf{Results.}
Table \ref{tab:overallperf} shows the performance of different methods on multi-page product search. Among all the methods, SCEM and SCRM3 perform better than all the other baselines without using short-term context, including their corresponding retrieval baseline, QEM, and QL respectively, and PROD which considers many additional features, showing the effectiveness of incorporating short-term context. 
In contrast to the effectiveness of short-term context, long-term context does not help much when combined with queries alone or together with short-term context. LCRM3 outperforms QL on all the datasets by a small margin; LCEM and LSCEM always perform worse than QEM and SCEM respectively when incorporating long-term context. 

QL performs similarly to RAND, which indicates that relevance captured by exact word matching is not the key concern in the rank lists of the production model. 
Most candidate products are consistent with the query intent but the final purchase depends on users' preference. Popularity, as an important factor that consumers will consider, can improve the performance upon QL. However, it is still worse than the production model most of the time.

\footnotetext{Due the confidentiality policy, the absolute value of each metric can not be revealed.}

We found that neural embedding methods are more effective than word-based baselines. QEM performs significantly better than QL, sometimes even better than PROD. When considering context, SCEM is much more effective than SCRM3. Neural embeddings capture not only semantic similarity but also co-occurrence of clicked and purchased items, which are more beneficial than exact word match for top retrieved items in product search. In addition, these embeddings also carry the popularity information since items purchased more will get more gradients during training. 
\section{Conclusion and Future Work}
\label{sec:conclusion}
We propose an end-to-end context-aware neural embedding model to represent various context dependency assumptions for predicting purchased items in multi-page product search. Our experimental results indicate that incorporating short-term context is more effective than using long-term context or not using context at all. It is also shown that our neural context-aware model performs better than the state-of-art word-based feedback models. 
Our work indicates that multi-page product search is a promising research topic. 
For future work, it would be better to evaluate our short-term context re-ranking model online, in an interactive setting as each result page can be re-ranked dynamically. 
Moreover, other information such as images and price can also be included to extract user preferences from their feedback.

\begin{acks}
This work was supported in part by the Center for Intelligent Information Retrieval and in part by NSF IIS-1715095. Any opinions, findings and conclusions or recommendations expressed in this material are those of the authors and do not necessarily reflect those of the sponsor.
\end{acks}

\bibliographystyle{ACM-Reference-Format}
\balance
\bibliography{reference}
\end{document}